\documentclass[journal=jpccck,manuscript=perspective]{achemso}

\usepackage[version=3]{mhchem} 
\usepackage{amsmath}
\usepackage{amsfonts}
\usepackage{amssymb}
\usepackage{bm}
\usepackage[T1]{fontenc}
\usepackage{xcolor}

\author{Ajay~Ram~Srimath~Kandada}
\email{srimatar@wfu.edu}
\affiliation[WFU]{Department of Physics and Center for Functional Materials, Wake Forest University, 1834 Wake Forest Road, Winston-Salem, NC~27109, United~States}

\author{Hao~Li}
\email{hli36@central.uh.edu}
\affiliation[UH]{Department of Chemistry, University of Houston, Houston, Texas~77204, United~States}

\author{Eric~R.~Bittner}
\email{ebittner@central.uh.edu}
\affiliation[UH]{Department of Chemistry, University of Houston, Houston, Texas~77204, United~States}

\author{Carlos~Silva-Acu\~na}
\email{carlos.silva@gatech.edu}
\affiliation[GT-chem]
{School of Chemistry and Biochemistry, Georgia Institute of Technology, 901 Atlantic Drive, Atlanta GA 30332, United~States}
\alsoaffiliation[GT-phys]
{School of Physics, Georgia Institute of Technology, 837 State Street, Atlanta GA 30332, United~States}
\alsoaffiliation[GT-MSE]
{School of Materials Science and Engineering, Georgia Institute of Technology, 771 Ferst Drive NW, Atlanta GA 30332, United~States}
\alsoaffiliation[CINVESTAV]{Departamento de F\'isica Aplicada, Centro de Investigaci\'on y de Estudios Avanzados del Instituto Polit\'ecnico Nacional, 97310 M\'erida, Yucat\'an, M\'exico}

\title[Exciton-polaron dephasing]
  {Homogeneous optical linewidths in hybrid Ruddlesden-Popper metal halides can \emph{only} be measured using nonlinear spectroscopy}

\begin{document}

\begin{tocentry}

\includegraphics[width=7.5cm]{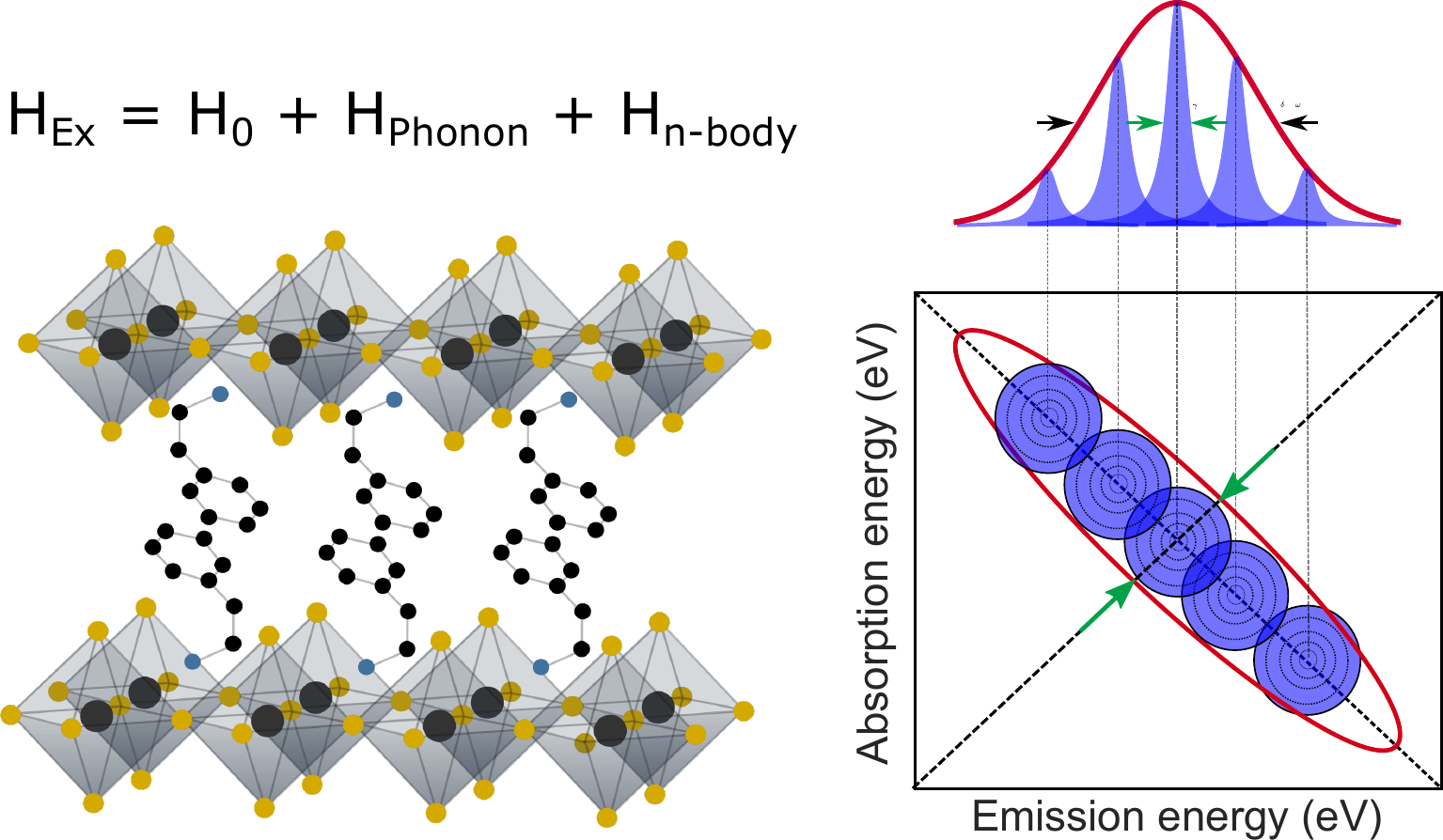}

\end{tocentry}

\begin{abstract}
The homogeneous photoluminescence spectral linewidth in semiconductors carries a wealth of information on the coupling of primary photoexcitations with their dynamic environment as well as between multi-particles. In the limit in which inhomogeneous broadening dominates the total optical linewidths, 
the inhomogeneous and homogeneous contributions can be rigorously separated by temperature-dependent 
steady-state photoluminescence spectroscopy. This is possible because the only temperature-dependent phenomenon is optical dephasing, which defines the homogeneous linewidth, since this process is mediated by scattering with phonons. However, 
if the homogeneous and inhomogeneous linewidths are comparable, as is the case in hybrid Ruddlesden-Popper metal halides, the temperature dependence of linear spectral measurement \emph{cannot} separate rigorously the homogeneous and inhomogeneous contributions to the total linewidth because the lineshape does \emph{not} contain purely Lorentzian components that can be isolated by varying the temperature. Furthermore, the inhomogeneous contribution to the steady-state photoluminescence lineshape is not necessarily temperature independent if driven by diffusion-limited processes, particularly if measured by photoluminescence. 
Nonlinear coherent optical spectroscopies, on the other hand, do permit separation of homogeneous and inhomogeneous line broadening contributions in all regimes of inhomogeneity. Consequently, these  
offer insights into the nature of many-body interactions 
that are entirely inaccessible to temperature-dependent linear spectroscopies. 
When applied to Ruddlesden-Popper metal halides, these techniques have indeed enabled us to quantitatively assess the exciton-phonon and exciton-exciton scattering mechanisms. 
Here, we will discuss our perspective on how the coherent lineshapes of Ruddlesden-Popper metal halides can be effectively rationalized within an exciton polaron framework.       

\end{abstract}

\newpage
\section{Introduction}

Two dimensional (2D) hybrid Ruddlesden-Popper metal halides, colloquially referred to as 2D hybrid perovskites, are self-assembled and solution-processed semiconductors that have multiple-quantum-well-like materials architectures and physical properties~\cite{mao2018two}. Current interest on these materials stems from their strongly excitonic properties that arise from electronic and dielectric confinement in two-dimensional layers of metal-halide octahedra that are separated by long organic cations such as phenethylammonium or n-butylammonium~\cite{katan2019quantum}. Strongly bound electron-hole pairs are thus observed in these materials and these are treated as Wannier-Mott excitons~\cite{even2014analysis} much like other two-dimensional semiconductors such as composite quantum wells and single-layer transition metal dichalchogenides. 

Given the immense interest in these materials for optoelectronic technologies, including in emerging quantum optoelectronics, where coherent properties of optical excitations play a primary role, quantifying the many-body interactions of excitons is extremely relevant. This includes measurement of exciton-phonon coupling, elastic multi-exciton scattering and biexciton binding energies. Linear optical spectroscopy, which measures either the absorption, photoexcitation, or photoluminescence of the sample, is widely employed for that purpose. For example, the temperature dependence of the photoluminescence linewidth is routinely analyzed to obtain the energy and coupling coefficients of phonons that scatter the excitons~\cite{gauthron2010optical,wehrenfennig2014homogeneous,wright2016electron,Neutzner2018exciton}. The density dependence of the PL spectra is considered to be an estimate of the biexciton binding energies~\cite{Ishihara1992}. It must be noted, however, that these methodologies were developed decades back mainly for epitaxially grown II-VI quantum wells and other similar systems,
which exhibit 
very low energetic disorder~\cite{lee1986luminescence, snoke2009solid}. Rigorous analysis of the PL lineshapes has severe limitations even in such systems due to many-body effects and non-trivial inhomogeneous contributions that arise in any ensemble measurement. Such contributions are even more dominant in solution-processed, self-assembled systems such as metal-halide hybrids. Moreover, existence of additional dynamic disorder effects from the organic-inorganic interactions~\cite{srimath2016photophysics}, strongly anharmonic lattice with substantial electron-phonon coupling puts metal-halide perovskites within a much more complex scenario. In this manuscript, we provide pertinent arguments to show that linear spectroscopy of such material systems, widely employed in the community, provides incomplete and often incorrect information. This may lead the over-simplification of the rather complex and unique photophysics simply because the data \emph{fits} to an established model. 

Coherent nonlinear spectroscopy~\cite{fuller2015experimental}, on the other hand, is a well suited technique to not only identify many-body interactions, but also to perceive the mechanistic subtleties of the scattering processes. Here, we provide an an extensive tutorial on the correlation between the coherent nonlinear spectral lineshapes and the many-body interactions and how the various bath interactions can be systematically decomposed and measured based on simple theoretical models. We will review two of our recent works --- Ref.~\citenum{thouin2019enhanced} and Ref.~\citenum{kandada2020stochastic} on a prototypical Ruddlesden-Popper metal halide, and we discuss the insights gained on the fundamentally unique nature of the excitons in these material systems.






\section{Exciton dephasing}

\begin{figure}[tbh]
    \centering
    \includegraphics[width=\textwidth]{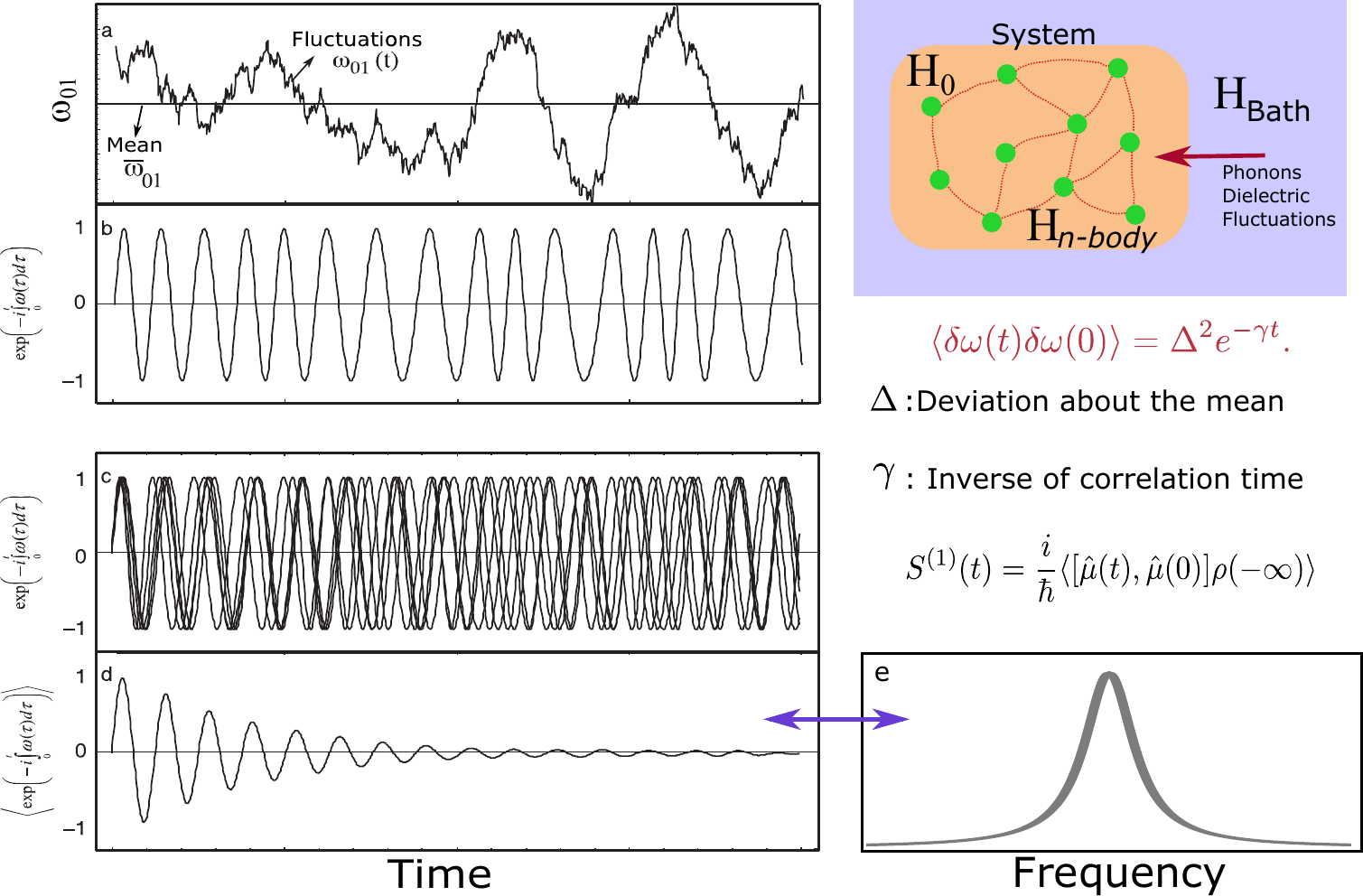}
    \caption{(a) Fluctuations in the exciton transition energy, $\omega_{01}$ around a mean value, driven by bath interactions, sketched on the right. (b) The value of the coherence term $\rho_{01}$ of a single particle excitation with a time varying frequency due to the fluctuations. (c) The coherence terms of a ensemble of excitations created in phase by the optical pulse, which evolve differently due to the random fluctuations. (d) The time average of the coherence term, which decays due to the dephasing process and whose Fourier transform is given by (e), which has a Lorentzian lineshape. Panels (a--d) reproduced with permission from Ref.~\citenum{hamm2011concepts}. Copyright 2011 Cambride University Press.}
    \label{fig:Kubo}
\end{figure}

In condensed matter, the spectral lineshape of an optical resonance encodes a wealth of information on the interactions of the system under investigation with its environment. The use of the spectral linewidth in atomic, molecular and semiconductor spectroscopies as a quantitative probe of system-bath interactions is extensively employed. The underlying physical principle behind this methodology was first formulated by Anderson~\cite{Anderson1954mathematical} and Kubo~\cite{kubo1969stochastic} within the framework of the stochastic scattering theory. In the single-particle Anderson-Kubo picture, the transition energy of the particle --- an atom, a molecule or an exciton in a crystal --- is modulated by the stochastic fluctuations in its environment. These fluctuations may be driven by scattering processes involving phonons (or local vibrations), dielectric fluctuations, or multi-particle collisions, see Fig.~\ref{fig:Kubo}. The result is an intrinsic time dependence of the transition frequency $\omega_{01}(t) = \omega_{01} + \delta\omega_{01}(t)$, where $\omega_{01}$ is the central frequency and $\delta\omega_{01}$ is the modulation such that $\langle \delta\omega_{01}(t) \rangle = 0$ (see Fig.~\ref{fig:Kubo}(a))~\cite{hamm2011concepts}.

In a simple two-level system, the transition probability from the ground state, $|0\rangle$ to the first excited state $|1\rangle$ is proportional to the off-diagonal element of the density matrix, $\rho_{01}$, which is also referred to as the coherence. In the absence of any perturbation to the system-bath Hamiltonian, the off-diagonal elements are zero. In other words, there is no projection of the ground state onto the excited state. An optical excitation induces a time-dependent component in the Hamiltonian: $H_{int}(t) = \vec{\mu}\cdot\mathbf{E}(t)$, where $\vec{\mu}$ is the transition dipole moment and $\mathbf{E}(t)$ is the electric field of the optical excitation pulse. The coherence term will accordingly oscillate as shown in Fig.~\ref{fig:Kubo}(b) following Equation~\ref{Eq:1}:

\begin{equation}
    \rho_{01}(t) = \rho_{01}(0) \exp \left( -i  \int_{0}^{t} \omega_{01}(\tau)\,d\tau \right ).
    \label{Eq:1}
\end{equation}
In an experiment, an ultrashort optical pulse excites an ensemble of particles with a well-defined phase coherence. However, with each of the particle's energy prone to statistically random fluctuations subject to the bath interactions, as shown in Fig.~\ref{fig:Kubo}(c), the measurement will perceive a time average of density matrix, $\langle \rho_{01}(t) \rangle$. The overall result is that the photoexcitations go out of phase exponentially over a period of time, referred to as the \textbf{dephasing time}, and the amplitude of the time averaged coherence term tends to zero as shown in Fig.~\ref{fig:Kubo}(d). The first-order optical response will be directly proportional to $\rho_{01}(t)$, the Fourier transform of which will be a Lorentzian lineshape, as in Fig.~\ref{fig:Kubo}(e). 
The prevalence of such a lineshape in an absorption or photoluminescence experiment, however, depends on the relative timescales of the fluctations. 

In order to be more quantitative about these timescales, we introduce a frequency autocorrelation function, $\langle \delta\omega(t) \delta\omega(0)\rangle$, which is a function of the frequency deviation about the mean, $\Delta$, and a certain correlation time for the fluctuations, $\tau_c = \gamma^{-1}$:   
\begin{equation}
    \langle \delta\omega(t) \delta\omega(0)\rangle = \Delta^2e^{-t/\tau_c}. 
    \label{Eq:2}
\end{equation}
The dephasing time is related to the correlation time, $T_2 = (\Delta^2\tau_c)^{-1}$~\cite{hamm2011concepts}. Note that this formulation has no underlying assumption concerning the nature of the environmental fluctuations. 
If $\Delta >> \gamma$, the fluctuations are relatively slow, enabling the system to sample a broad inhomogeneous distribution of environmental conditions. This condition does not result in the time-domain response shown in Fig.~\ref{fig:Kubo}(d) and accordingly, the absorption/emission spectrum takes a Gaussian lineshape whose width has no dependence on the correlation time, but is rather limited by the width of the inhomogeneous distribution sampled, and accordingly the lineshape does \emph{not} carry any information on the bath interactions. On the other hand, if the fluctuations are rapid enough, $\Delta << \gamma$, in the motional narrowing limit, the linewidth becomes increasingly narrow and assumes a Lorentzian form where the \textbf{homogeneous} linewidth, $\gamma = \tau_c^{-1}$ is now directly related to the dephasing time and thus quantifies the system-bath interactions. An intermediate regime with comparable homogeneous and inhomegeneous contributions to the linewidth is possible in systems that are subjected to static and dynamic disorder. We note that highly crystalline systems can also fall into this category subject to the inevitable disorder from the fabrication.~\cite{siemens2010resonance} In such cases, the optical lineshape has a much more complex behaviour with the linewidth which is not directly correlated with he dephasing rate albeit being dependent on it. \\

\subsection{Thermal dephasing probes}

Of the various sources of environmental stochastic fluctuations, 
thermal excitation is a dominant one. Thermally populated vibrations, dielectric fluctuations and, in the case of crystals, phonons, scatter electronic excitations. This can be perceived as an increasing linewidth at higher temperature and accordingly it is common to analyse the temperature dependence of the linear spectral linewidths measured via absorption or photoluminescence, to quantitatively estimate the electron-phonon coupling parameters. 
The homogeneous PL linewidth ($\gamma$) is routinely modelled assuming scattering of electrons with polar optical phonons and with long-wavelength acoustic phonons:  
\begin{equation}
    \gamma(T) = \gamma_0 + \alpha_{ac}T + \alpha_{opt}N_{opt},
    \label{Eq:3}
\end{equation}
where $\gamma_0$ is the temperature independent, intrinsic linewidth, while $\alpha_{ac}$ and $\alpha_{opt}$ are the electron-acoustic phonon and electron-optical phonon coupling coefficients respectively. The number of the optical phonons with energy $E_{LO}$ is $N_{opt}$, which is given by the Bose-Einstein distribution function, $1/\left[ {\exp \left (E_{LO}/k_BT \right )-1}\right]$. It is worthwhile to note that this model was developed primarily in the context of material systems fabricated with a low degree of disorder, such as epitaxial semiconductor quantum wells and quantum dots. This is evident from the reported linewidths that are well below 10\,meV (less than the relevant phonon energies) and the observed near-perfect Lorentzian spectral lineshapes. However, there are still contributions from inhomegenous broadening effects due measurable and unavoidable static disorder, that limit application of Eq.~\ref{Eq:3} even in these model material systems. While these contributions can be reduced via single-particle spectroscopy, they are inevitable in an ensemble measurement. In fact, the need to go beyond linear spectroscopies to measure true homogenous linewidths has been emphasised in several published works~\cite{liu2021toward, moody2015intrinsic, siemens2010resonance}. 

Metal-halide perovskites and their derivatives exhibit PL linewidths that are factors larger in magnitude than the relevant phonon energies with lineshapes that are distinctly 
not Lorentzian~\cite{wright2016electron, gauthron2010optical}. This may be attributed fundamentally to their self-assembly process from solution, which is bound to generate substantial static morphological disorder. In addition, 
dynamic disorder is ubiquitous in the hybrid 
metal-halide lattice due to the independent motion of the organic cation. These sources of disorder inevitably increase the deviation in the frequency fluctuations, thus forcing the system to be in the limit where the lineshape is a inhomogeneous distribution of homogeneously broadened transitions, see Fig.~\ref{fig:2Dlineshape}(a). The contribution of \emph{static} and \emph{dynamic} disorder is therefore comparable in these systems. In this condition, the PL linewidth is 
neither dominated by homogeneous or inhomogeneous mechanisms, and Eq.~\ref{Eq:3} is not valid. In other words, the temperature dependence of the linewidth is not uniquely attributed to scattering with optical or acoustic phonons. Accordingly, the values of the coupling coefficients and phonon energies obtained from such an analysis, particularly in metal halide perovskites, are at best upper bound estimates, and a \emph{direct} measurement of the homogeneous lineshape is necessary. 

This model is yet widely used to derive very impactful conclusions on the photophysics of three dimensional metal halide perovskites~\cite{wright2016electron, wehrenfennig2014homogeneous}, hybrid layered metal halides (see Refs.~\citenum{Neutzner2018exciton, ni2017real, long2019exciton, straus2019longer, zhang2018optical, esmaielpour2020role} for a few examples) and nanostructured derivatives (see for examples Refs~\citenum{peng2020suppressing, lao2019anomalous,saran2017giant,ghosh2018phonon, rubino2020disentangling}). While some of these published works qualitatively captured the nature of carrier-phonon and exciton-phonon interactions, none of the reported linear spectroscopies were able to provide quantitative estimates of the coupling coefficients by the arguments presented above. The insufficiency in this methodology is also evident in the wide variance in the reported phonon energies and coupling coefficients, with some of them estimating the relevant phonon energies to be above 50\,meV~\cite{zhang2018optical, straus2019longer}, factors above the rigorously estimated LO phonon energies of less than 10\,meV in these material systems~\cite{yaffe2017local, Thouin2018stable}.   

One might argue that the inhomogeneous contributions to the linewidth can be modelled as a temperature independent offset $\gamma_0$ in Eq.~\ref{Eq:3}. However, the photoexcitation can sample a broad range of site energies by diffusion processes that are temperature dependent and that happen within its lifetime. 
Thus, luminescence spectroscopy in particular, and any linear spectroscopy in general, is incapable of separating the fast dephasing dynamics and slow evolution of the population within the inhomegeneous energetic landscapes. 
Therefore, spectroscopic techniques that can \emph{directly} distinguish between homogeeous and inhomogeneous contributions to the total linewidth are required. 
Coherent multi-dimensional spectroscopy thus offers a viable alternative here, as demonstrated recently in Refs.~\citenum{thouin2019enhanced, liu2021toward, yu2021exciton}. 

\subsection{Two-dimensional coherent optical spectroscopy}

\begin{figure}
    \centering
    \includegraphics[width=0.9\textwidth]{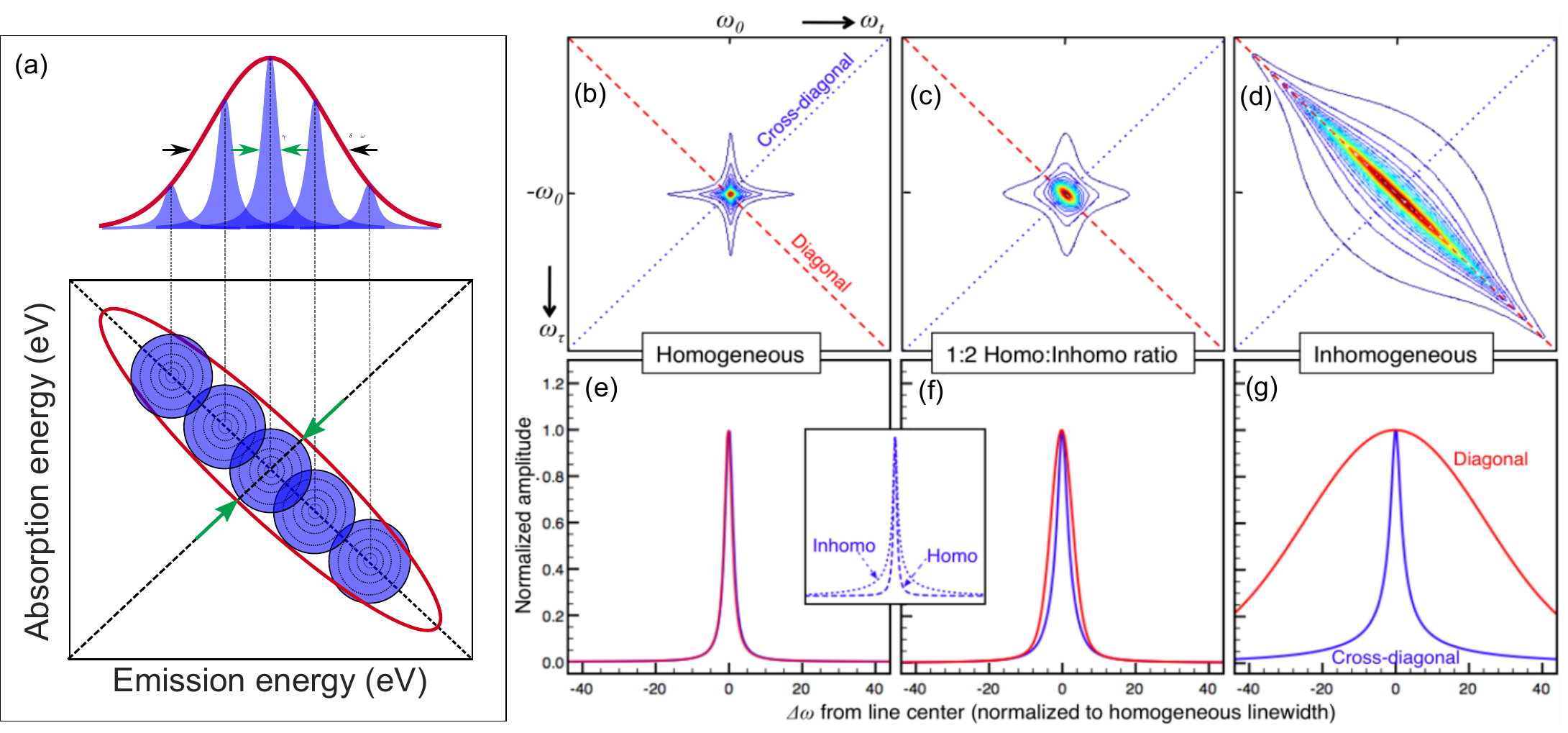}
    \caption{(a) Schematic representation of an inhomogeneously broadened Gaussian linear lineshape, with a distribution of Lorentzian homogeneously boradened linshapes buried underneath it. In this scenario, the 2D coherent lineshape reflects the inhomogeneous, Gaussian spectrum along the diagonal spectral cut, and the homogeneous, Lorentzian spectrum along the anti-diagonal. The 2D coherent lineshape is depicted in conditions of (b) $\gamma \gg \Delta$ (homogeneous broadening), (c) $\gamma \sim \Delta$ (moderate inhomogeneity), and (d) $\gamma \ll \Delta$ (inhomogeneous broadening), along with corresponding diagonal and anti-diagonal cuts in (e), (f), and (g), respectively. Panels (b)--(g) reproduced with permission from Ref.~\citenum{siemens2010resonance}, Copyright 2010 The Optical Society.}
    \label{fig:2Dlineshape}
\end{figure}

In a two-dimensional coherent spectroscopy experiment, the non-linear optical response of the sample is measured as a function of two independent energy (or frequency) variables~\cite{tokmakoff2000two}. The experiment itself involves photo-excitation of the sample with three phase-stabilized, ultrashort (typically $\lesssim 25$\,fs) optical pulses. The excitation geometry, commonly know as the boxCARS geometry, is such that each of the three pulses are at the vertices of a square while being focused onto the sample. The three optical pulse interaction induces a time-varying third-order polarization in the material that emits coherent radiation carrying the nonlinear material response. Due to the specific excitation geometry and the resulting phase-matching and time-ordering conditions, this signal is emitted along the fourth vertex of the square in the boxCARS geometry. The phase and amplitude of the signal is then measured by interfering it with another low intensity optical pulse, a local oscillator with known phase and amplitude. The energy of the detected signal constitutes one of the axes of the two-dimensional spectrum and referred to as the \textbf{emission energy}, see Fig.~\ref{fig:abs_rephasing}(b). Note that this is not the energy of the luminescence emitted by the photo-excited state via population relaxation process, but it is the energy of the coherent signal emitted by the time-evolving coherent nonlinear polarization. 

The second axis in the 2D correlation map, referred to as the \textbf{absorption energy} is the energy at which the photo-excited population is created. This axis is constructed as the Fourier transform of the signal dynamics obtained by scanning the delay between the first two excitation pulses. Note that in a three-pulse excitation scheme, the first pulse creates a coherence like in Eq.~\ref{Eq:1} and the second pulse then projects the coherence onto a population and the phase evolution of the signal in between the first two pulses corresponds to the absorption event.

Fig.~\ref{fig:2Dlineshape}(a) summarizes, albeit in a qualitative way, how a 2D coherent spectrum can separate inhomogeneous and homegeneous contributions, which manifest as diagonal and anti-diagonal linewidths respectively. In the limit of motional narrowing ($\Delta \ll \gamma$), however, there is no distinction between diagonal and anti-diagonal lineshapes, and the 2D spectrum is symmetric and \emph{star-like} as shown in Fig.~\ref{fig:2Dlineshape}(b). In the other extreme limit of inhomogeneous broadening ($\Delta \gg \gamma$), the difference between the diagonal and anti-diagonal linewidths is evident (see Fig.~\ref{fig:2Dlineshape}(d)) and $\gamma$ can be estimated from the latter. In the intermediate regime ($\gamma \sim \Delta$), like in the case of Ruddlesden Popper metal-halides and other metal halide perovskite derivatives~\cite{Neutzner2018exciton,Thouin2018stable,thouin2019enhanced,kandada2020stochastic}, diagonal and anti-diagonal widths 
depend on both $\gamma$ and $\Delta$, and both linshapes become coupled as shown in Fig.~\ref{fig:2Dlineshape}(c). 

In a such a scenario, a global fitting routine based on analytical expressions for the lineshapes derived by Cundiff and coworkers~\cite{siemens2010resonance} must be employed to quantitatively estimate the homogeneous linewidth. There are additional experimental considerations based on the sequence of the optical pulses, phase-matching conditions and appropriate population waiting times that enable true estimation of the homogeneous linewidth. While more extensive treatments can be found elsewhere~\cite{hamm2011concepts, tokmakoff2000two}, it can be shown that the absolute value of the \emph{rephasing} spectrum that is collected at (or close to) the \emph{zero} population waiting time is the correct spectrum to be used in the analysis. The rephasing spectrum refers to the experimental condition in which the first pulse interacting with the sample is the one which is at the diagonally opposite vertex to the detecting local oscillator pulse in the boxCARS geometry. The rephasing signal is also referred to as the \emph{photon echo} generated through a four wave mixing mechanism. We note that while there are other relatively simple experimental implementations of the 2D coherent spectroscopy~\cite{fuller2015experimental}, some of them are not capable of measuring reliable homogeneous linewidths due to their inability to isolate the coherent excitation pathways that give the  appropriate total nonlinear response. It is important to underline that only the rephasing coherent pathway can quantitatively resolve the homogeneous contributions to the total coherent lineshape under general conditions. 


\subsection{Measurement of homogeneous linewidths in \ce{(PEA)2PbI4}}

\begin{figure}
    \centering
    \includegraphics[width=0.9\textwidth]{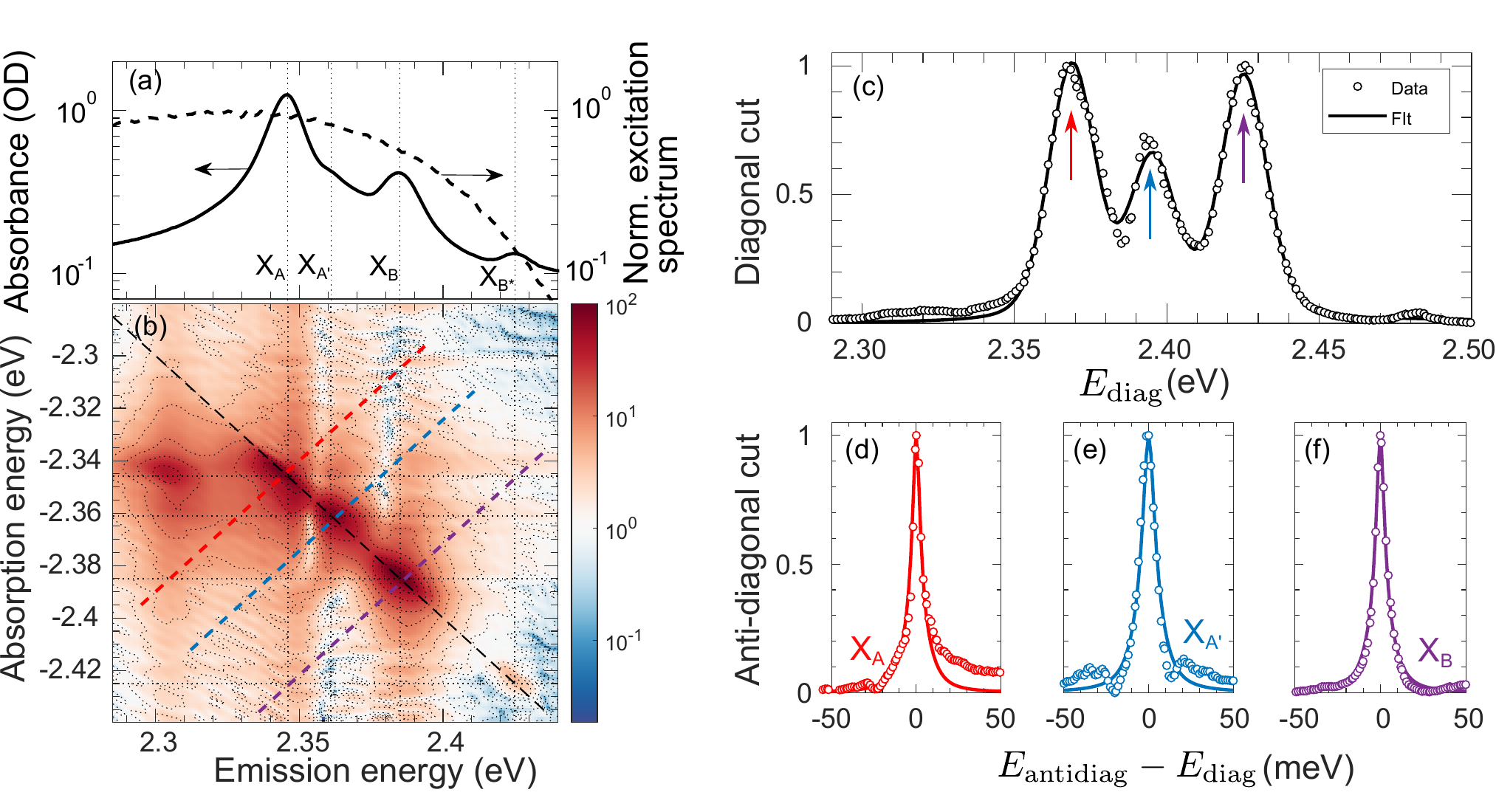}
    \caption{Linear and two-dimensional coherent spectroscopy of \ce{(PEA)2PbI4} at 5\,K. (a) Absorption spectrum of \ce{(PEA)2PbI4} measured at 5\,K (black line) and normalized spectrum of the pulses used in 2D coherent excitation spectroscopy measurements (dashed line). Both scales are logarithmic. Dotted lines indicate the energy of excitons A, A$^{\prime}$, B, and B$^*$, respectively, with increasing energy. (b) Absolute value of the 2D coherent rephasing spectrum of \ce{(PEA)2PbI4} measured at 5\,K with a pulse fluence of 40\,nJ/cm$^2$ and a pump-probe delay of 20\,fs. The color scale is logarithmic. Dotted lines indicate the energies of the aforementioned features. The paths of the diagonal cut (black dashed line) and antidiagonal cuts at the diagonal energy of excitons A, A$^{\prime}$, and B (red, blue, and purple dashed lines, respectively) are also shown. Figure reproduced with permission from ref.~\citenum{thouin2019enhanced}. Copyright 2019 American Physical Society. }
    \label{fig:abs_rephasing}
\end{figure}

A representative zero-time rephasing 2D spectrum of a prototypical Ruddlesden Popper metal halide taken at 5\,K is shown in Fig.~\ref{fig:abs_rephasing}(b). It must be noted that the spectral response at zero population time may be affected by the presence of coherent artefacts and pulse-ordering ambiguities. We exclude such contributions to the measured spectrum, based on the consistent time evolution of the spectral lineshape, which we discussed in Refs.~\citenum{thouin2019enhanced, kandada2020stochastic} and we will come back to these dynamics later in this manuscript. We can identify three features along the diagonal corresponding to the three resonances in the linear spectrum in Fig.~\ref{fig:abs_rephasing}(a) that correspond to the distinct excitonic states~\cite{Neutzner2018exciton,thouin2019phonon,SrimathKandada2020}. The 2D lineshape clearly represents the intermediate condition of Fig.~\ref{fig:2Dlineshape}(c) and we performed a global fit of the diagonal (Fig.~\ref{fig:abs_rephasing}(c)) and anti-diagonal cuts (Fig.~\ref{fig:abs_rephasing}(d)--(f)). We estimated comparable homogeneous and inhomogenous linewidths for this particular spectrum to be approximately 2.3 and 6.5\,meV, respectively~\cite{thouin2019enhanced}.

\begin{figure}
    \centering
    \includegraphics[width=0.9\textwidth]{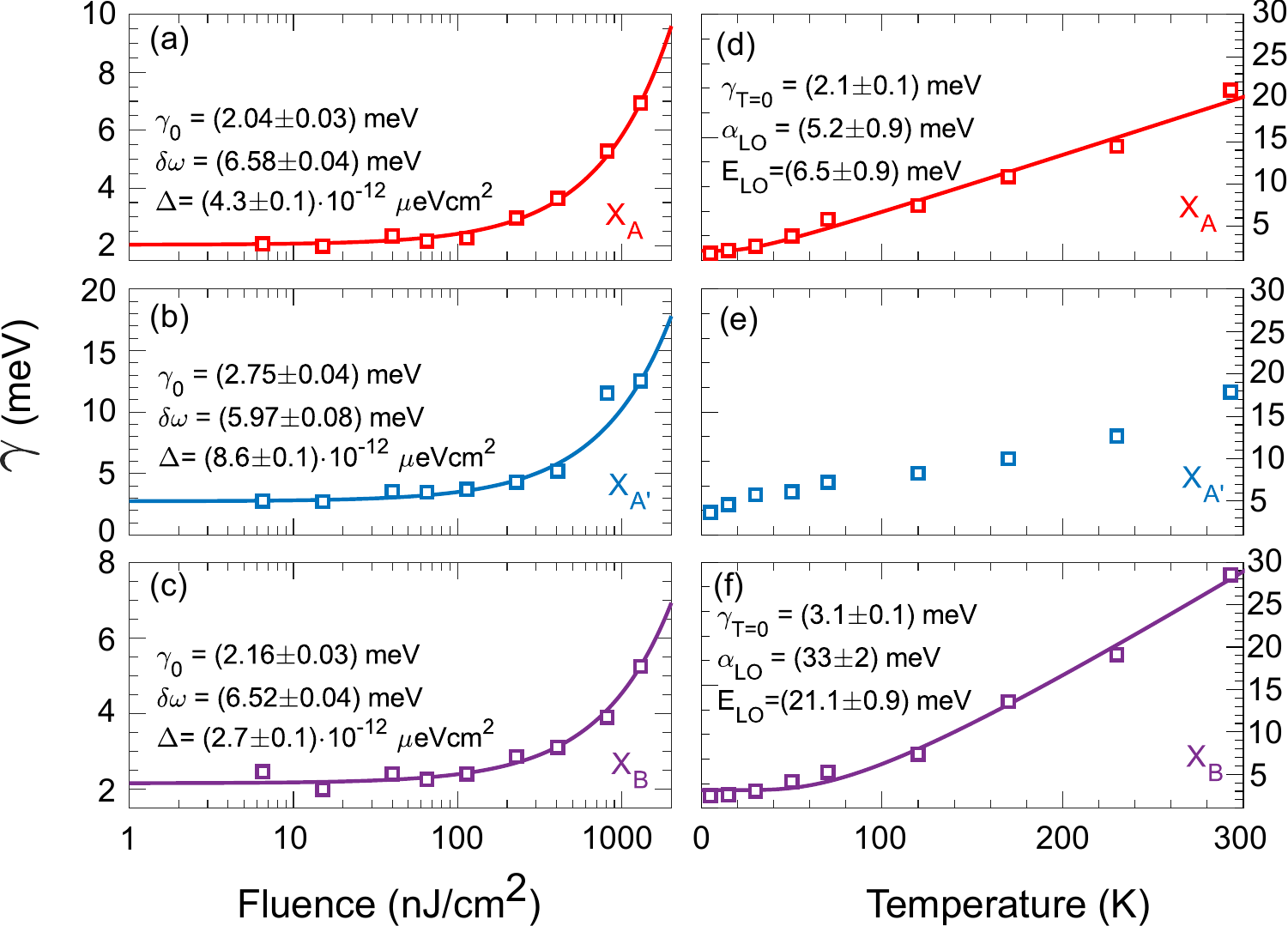}
    \caption{Fluence and temperature dependence of the exciton dephasing rates. Dephasing parameters $\gamma$ of excitons A, A$^{\prime}$, and B [panels (a) and (d), (b) and (e), and (c) and (f), respectively) obtained from the simultaneous fitting of diagonal and antidiagonal cuts, plotted as a function of excitation fluence [panels (a)–(c)] or temperature [panels (d)–(f)]. Squares represent the experimental linewidths and lines are the best fit to the relevant model described in the main text. Error bars on the data are contained within the markers. For panels (a)–(c), the sample temperature is maintained at 5\,K while the excitation fluence is kept at 50\,nJ/cm$^2$ for measurements presented in panels (d)–(f). Figure reproduced with permission from ref.~\citenum{thouin2019enhanced}. Copyright 2019 American Physical Society.}
    \label{fig:density-temp}
\end{figure}

The homogeneous linewidth is strongly dependent both on the excitation fluence and the sample temperature owing to the the many-body exciton interactions, as shown in Fig.~\ref{fig:density-temp}. The linewidth increases linearly with increasing excitation density owing to exciton-exciton scattering~\cite{thouin2019enhanced}, also referred to as the excitation induced dephasing (EID), with the slope of the curve quantifying the strength of such interactions~\cite{moody2015intrinsic}. It is notable that the strength of the inter-exciton interactions is 
distinct for each of the observed resonances. We will discuss the nature of EID later in this perspective while noting the evolution of the 2D lineshape with the population waiting time. 

The thermal dephasing of the excitons, on the other hand, can be perceived by the temperature dependence of the linewidth. Having separated homogeneous and inhomogeneous contributions rigorously, Eq.~\ref{Eq:3} can now be reliably used to analyse this dataset. We obtain good fits with the LO phonon scattering model for each of the observed resonances within the spectral finestructre at the exciton energy labelled $X_{A}$ and $X_{B}$ (see the last section of this article for more discussion) and we estimate the energy of the phonon(s) participating in the elastic scattering process and the value of the exciton-phonon coupling coefficient. The thermal dephasing of exciton $X_{A}$ indicates that it is scattered by a LO phonon at 6.5\,meV~\cite{thouin2019enhanced}, which we had previously identified through impulsive vibrational spectroscopy~\cite{thouin2019phonon}. In the case of $X_{B}$ however, the phonon energy is estimated to be 21\,meV with substantially larger exciton-phonon coupling constant~\cite{thouin2019enhanced}. 

The phonon energy corresponding to $X_{B}$ is closer to the predictions from PL linewidth analysis on similar material systems (see Refs.~\citenum{Neutzner2018exciton,gauthron2010optical} for the analysis on \ce{(PEA)2PbI4)}. We note, however, that there are no LO phonon modes involving the metal-halide lattice motion at energies above 10\,meV as predicted by DFT theory and resonance Raman experiments~\cite{thouin2019phonon}. There are certainly modes that correspond to the motion of the organic cation, specifically the $\pi-\pi$ motion of the phenyl group that can be expected to have similar energies~\cite{dragomir2018lattice,urban2020revealing}. The mechanism that can drive the scattering of excitons that are confined within the two-dimensional metal-halide layer with uncorrelated, localized vibrations of the organic cation is, however, unclear if not physically implausible. We consider that the LO phonon scattering model is thus insufficient to explain the thermal dephasing of excitons in Ruddlesden-Popper metal halides. This is further substantiated by the failure of the model to reproduce the thermal trend of $X_{A^{\prime}}$. We believe that the exciton polaron framework can address this discrepancy, 
since thermal dephasing of exciton polarons will intrinsically involve the dephasing of the phonon component of the quasi-particle through the anharmonic lattice which is not captured by Eq.~\ref{Eq:3}~\cite{rojasgatjens2021}. We highlight that such intricacies in the scattering mechanisms cannot be reliably derived by simply performing temperature-dependent photoluminescence measurements.

\subsection{Excitation Induced Dephasing}


The fluence dependence of the linewidths over three orders of magnitude in the excitation densities, as shown in Fig.~\ref{fig:density-temp}, is indicative of the dominance of excitation-induced dephasing in Ruddlesden Popper metal halides. As noted several times in this manuscript, such a dependence can never be observed in a linear PL measurement, which highlights the importance of nonlinear coherent spectroscopy. The optical response function of semiconductor materials is a complex quantity, and coherent spectroscopy enables the separation of the real and imaginary components of the response function via the simultaneous measurement of the amplitude and phase~\cite{hamm2011concepts}. 

The linear optical absorption spectrum, measured via the transmission amplitude, corresponds to the imaginary component of the complex 
permittivity. A resonance in the absorption spectrum is accordingly manifested as a dispersive, first-derivative-like lineshape in the real part of the linear spectrum, following the Kramers-Kronig relations. Nonlinear coherent spectroscopies can simultaneously provide both the real and imaginary components of the optical response, enabling such absorptive-dispersive lineshapes to be identified in the nonlinear spectrum without the need of Kramers-Kronig analysis. 
When the coherent response is phased correctly, and in the absence of many-body interactions, the real part of the 2D spectrum has a symmetric absorptive lineshape, while the imaginary component is dispersive~\cite{shacklette2003nonperturbative}.

In the presence of many-body interactions, which scramble the relative phase of the nonlinear response, the real part of the rephasing spectrum (and the non-rephasing spectrum) acquires a dispersive lineshape while the imaginary component looks distinctly absorptive. Several earlier works established such dispersive 2D lineshape as an indicator of the elastic exciton-exciton scattering through a phenomenological model based on optical Bloch equations~\cite{shacklette2002role, shacklette2003nonperturbative, Li2006}. The model considers two distinct consequences of the Coulomb-mediated interactions between excitons --- (i) excitation induced shift in the exciton energy and (ii) excitation induced dephasing which broadens the transition, both of which generate population density dependent terms in the time evolution of the coherence function (Eq.~\ref{Eq:1}). Notably, this model assumes a population of excitons as the bath, which remains stationary  within the dephasing time. Accordingly, while the 2D spectral intensity reduces and the lineshape elongates along the diagonal due to spectral diffusion driven by energy migration, the dispersive component of the lineshape remains intact during the population waiting time ($\tau_p$).

\begin{figure}[tbh]
    \centering
    \includegraphics[width=\textwidth]{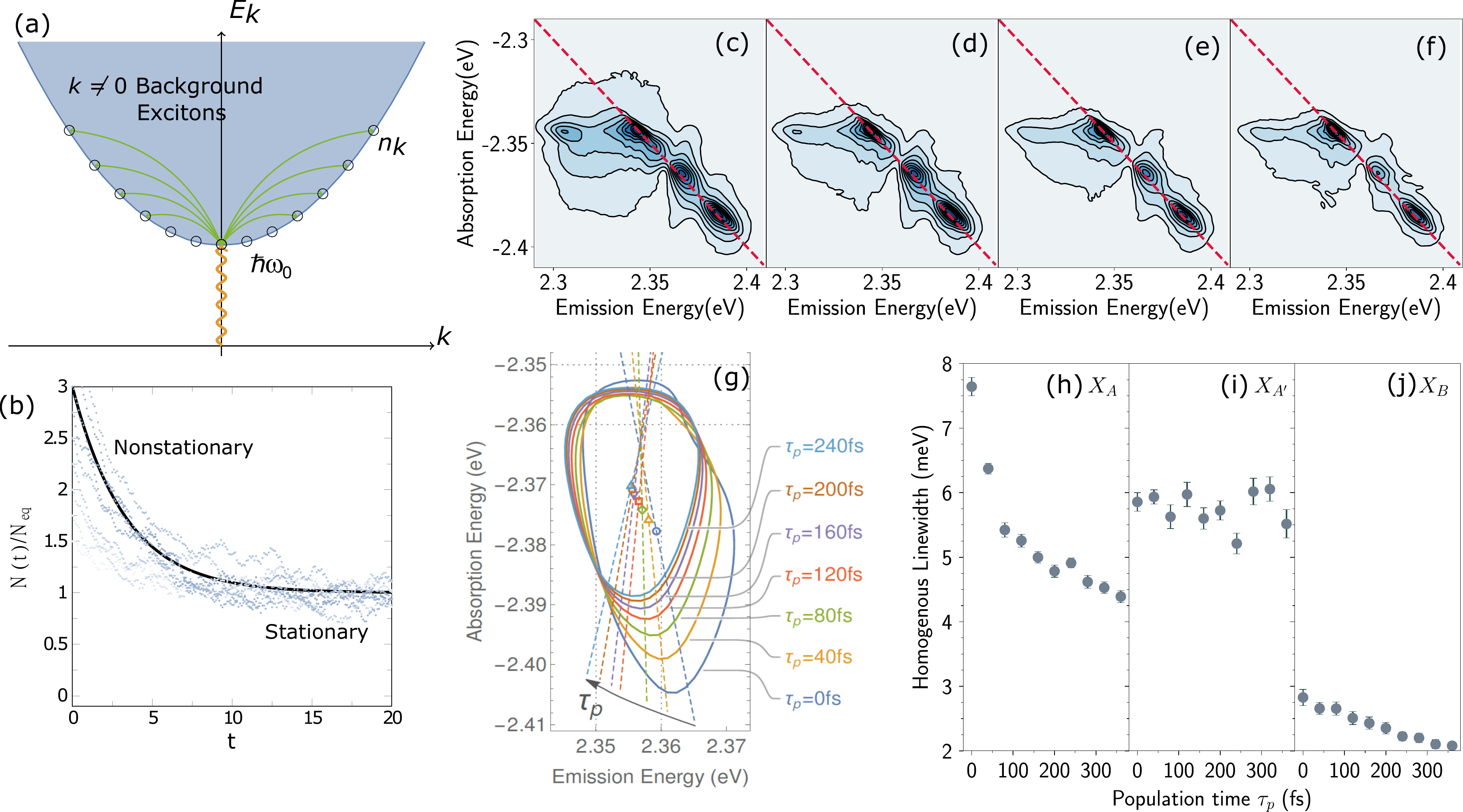}
    \caption{(a) Schematic representation of optical absorption of excitons and exciton-exciton scattering with a background population, where the dispersion relation is in the exciton representation and $\vec{k} = \vec{k_e} + \vec{k_h}$ is the exciton wavevector. (b) Time evolution of the non-stationary background population, which reaches a stationary condition asymptotically. (c)--(f) The absolute values of the experimental rephasing spectrum of \ce{(PEA)2PbI4} at population waiting times of 0\,fs, 80\,fs, 160\,fs and 240\,fs respectively, where clear narrowing of the lineshape can be identified. This can also be seen more quantitatively as decreasing homogeneous linewidth with the population waiting time for (h) $X_A$ (i) $X_{A'}$ and (j) $X_B$. (g) Calculated coherent lineshape contour based on model represented in (a) which reproduces the narrowing of the linewidth. Figure reproduced with permission from ref.~\citenum{kandada2020stochastic}. Copyright 2020 American Institute of Physics. }
    \label{fig:EID}
\end{figure}

In the case of \ce{(PEA)2PbI4}, we observed a very evident dispersive lineshape in the real part of the rephasing spectrum at each of the three exciton resonances~\cite{kandada2020stochastic}, as expected from a system that is subject to many-body interactions (see also Fig.~\ref{fig:density-temp}(a)--(c)). The lineshape, however, exhibited a very strong population-time dependence with the dispersive lineshape evolving into the absorptive lineshape within 500\,fs. A correlated change can also be observed in the imaginary component of the response, which progressively acquired dispersive component. These experimental observations indicate that the many-body exciton-exciton interactions are being quenched in a timescale where there is negligible loss in the population (photo-generated excitons do not recombine to the ground state withing 500\,fs).

The time-evolving many-body interactions can also be observed through the absolute values ($\sqrt{(\Re)^2 + (\Im)^2}$) of the rephasing spectrum shown in Fig.~\ref{fig:EID}(c)--(f). The narrowing of the spectral lineshape is evident in this dataset, which can be further analysed to obtain the time dependent homogeneous linewidths for the excitonic transitions - $X_A$, $X_A^{\prime}$ and $X_B$, shown in Fig.~\ref{fig:EID}(h)--(j). Intriguingly, the reduction in the homogeneous linewidth is very different for each of the resonances. To understand this, we must first understand the microscopic origin of the time-dependent linewidth.

In Refs.~\citenum{kandada2020stochastic} and \citenum{Li2020stochastic}, we developed a first-principles, many-body theoretical model of interacting excitons employing a quantum stochastic approach, which provided a theoretical framework in which these experimental observations can be rationalized. Our approach was similar to the Anderson-Kubo (AK) model of system-bath fluctuations, which we elaborated upon earlier in this manuscript. The notable difference between the AK model and our treatment lies in the assumption on the nature of the stochastic bath fluctuations. Conventionally, the bath fluctuations are considered to be around a stationary state and the energy fluctuation average around a time-independent mean value, see Fig.~\ref{fig:Kubo}(a). But with the time evolving coherence, we considered a non-stationary bath population whose relaxation is described by a stochastic equation, see Fig.~\ref{fig:EID}(b).

The physical picture behind our model is shown in Fig.~\ref{fig:EID}(a): at $t = 0$, a sequence of broadband femtosecond pulses photo-excite a non-stationary population of excitons at non-zero momenta in the exciton phase space in addition to the coherent population at $k=0$. The coherent population is eventually detected via the nonlinear measurement through a very specific coherent excitation pathway and it is the \emph{system} whose spectral response is being measured. However, the energy of the coherent excitons is modulated by the Coulomb mediated interactions with the noisy bath of the incoherent exciton population at $k \ne 0$ with the fluctuation, $\Delta$ proportional to the background population density, $N(t)$). 
The fluctuations in $\omega_{01}(t)$ in Eq.~\ref{Eq:1} can accordingly be derived from the fluctuations in the background population density $N(t)$ that scatter the coherent exciton population via a many-body interaction potential. Under the Born approximation and for a finite range potential, this interaction can be described in terms of the s-wave scattering length ($a$) and the exciton effective mass ($\mu$) as $V_0 = 4\pi \hbar^2 a/\mu$. $N(t)$ itself is described by Ornstein-Uhlenbeck process (damped Brownian motion) that results in a the mean of the population to be $\langle N(t) \rangle = e^{-\gamma^*t} \langle N(t=0) \rangle$, where $\gamma^*$ is the background relaxation rate, as shown in Fig.~\ref{fig:EID}(b).


Following this photophysical scenario, the optical response of function which depends on the mean of the time dependent coherence function in Eq.~\ref{Eq:1} can be shown to be proportional to~\cite{Li2020stochastic}:
\begin{equation}
   \left \langle \exp \left [ -i2V_0 \int_{0}^t N(\tau) d\tau \right] \right \rangle = \exp \left[ i2V_0g_1(t)\right] \exp \left[ - 2V_0^2 g_2(t) \right],
   \label{Eq:4}
\end{equation}
where $g_1(t)$ and $g_2(t)$ are the time averaged mean and the covariance of the background population fluctuations. In simple words, they quantify the time evolution of the background population and its fluctuations. Eq.~\ref{Eq:4} shows that the coherence function and accordingly the optical response oscillates at a frequency which is renormalized by $g_1(t)$ and results in the excitation induced shift to blue in the exciton energy. The coherence also decays at a rate determined by $g_2(t)$, faster than the intrinsic dephasing rate and which is density dependent and thus leads to the excitation induced dephasing and the phase scrambling that gives rise to the dispersive lineshape. 

The experimental results on \ce{(PEA)2PbI4}, shown in Fig.~\ref{fig:EID} reveal more dominant contribution from line narrowing and phase scrambling than from the excitation induced shifts. This suggests a greater role from the covariance in the population fluctuations than the overall decay in the time averaged mean. The dynamical narrowing of the rephasing spectral lineshape can be reproduced within this framework as shown in Fig.~\ref{fig:EID}(g). The theory also predicts substantial asymmetry in the lineshape along the absorption energy axis which is clearly absent in the experimental spectrum. We attribute this again to the reduced contribution from $g_1(t)$. Importantly, the critical physical mechanism is not just the overall decay of the background population but the stochastic evolution of correlations in the density fluctuations which in itself may be driven by the lattice induced polaronic effects. Coincidentally, the time evolution of the coherent nonlinear lineshape (albeit the total currelation spectrum, not resolved into the rephasing component) has also been recently observed by Kambhampati and co-workers in lead-halide nanocrystals, and was suggested to be a signature of the formation of exciton polaron~\cite{sonnichsen2021polaronic}.

\section{Perspective}\label{sec:Pers}

The spectral finestructure in the optical spectra of hybrid Ruddlesden Popper metal halides has largely been suggested to be composed of phonon replicas of a single excitonic state. We have argued that this may not be a sufficient explanation of the spectral structure based on a series of experimental observations. Two dominant resonances, $X_A$ and $X_B$ can be identified within the spectral structure separated by a characteristic energy of 35$\pm$5\,meV. The spectral lineshape can be reproduced with a empirically modified Elliott's formula, also with the Huang-Rhyss contributions, but not with phonons at 35\,meV~\cite{Neutzner2018exciton, passarelli2020tunable}. In fact, we do not observe phonons at that energy that are strongly coupled to electronic excitations in the Resonance resonant impulsive stimulated Raman scattering measurement. Instead, as we elaborated in Ref.~\citenum{thouin2019phonon}, we observe that the impulsive stimulated Raman spectra, obtained by exciting resonantly and independently the two resonances, are very distinct, implying that the $X_A$ and $X_B$ are dressed by different phonons. The recombination dynamics of observed via time resolved photoluminescence and transient absorption spectroscopies and their temperature dependence indicate non-adiabatic mixing of the excitonic states assisted by the lattice phonons, as elaborated in Ref.~\citenum{thouin2019polaron}. The energy and sign of the biexciton binding energies of the two resonances estimated via multi-quantum two-dimensional spectroscopy are very different for the two resonances are reported in Ref.~\citenum{Thouin2018stable}. These observations strongly suggest that these resonances correspond to distinct electronic states and are not part of single electronic manifold. There are a few recent and excellent mangentoabsorption data~\cite{urban2020revealing} which may substantiate the vibronic assignment and we will provide our take on those observations in a separate article. 

Here, we instead highlight the lessons learnt from the dephasing dynamics observed and analysed by us in \ce{(PEA)2PbI4}. If $X_A$ and $X_B$ were indeed phonon replicas, then excitation of $X_B$ will photo-generate a phonon of 40\,meV energy in addition to the excitonic state similar to that of $X_A$. Given that the phonon energy is an order of magnitude lower than the exciton binding energy and there is no coherent interaction/superposition of the exciton and phonon states in this picture, one would expect that the probability of many-body elastic scattering of $X_A$ excitons will be similar, if not identical to that of $X_B$ excitons. The experimental observation in Fig.~\ref{fig:density-temp} indicates the contrary with evident distinctness in the dephasing rates of $X_A$ and $X_B$ and their temperature and fluence dependence. We already discussed the peculiarity in the thermal dephasing of these excitons and the inadequacy in the currently employed LO phonon scattering models in this context earlier in this manuscript. We reiterate the importance of phonon-phonon interactions to properly account for the thermal dephasing mechanism, see the discussion in Ref.~\citenum{rojasgatjens2021}.

The rate of increase in the dephasing rate with increasing excitation density, albeit being in the similar order of magnitude, is considerably different for each of the resonances. The dispersive lineshape and time evolution of the homogeneous linewidths as shown in Fig.~\ref{fig:EID} are also very evidently different for each of the resonances. Considering the stochastic scattering of the coherent exciton population with the $k\ne 0$ excitons as the dominant dephasing mechanism, under the Born approximation, we can deduce that that the many-body interaction potential $V_0$ is fundamentally different for $X_A$, $X_A'$ and $X_B$. This suggests that the characteristic exciton parameters such as the s-wave scattering lengths (linked with the exciton size and its charge-transfer character) and the exciton effective mass are measurably different. These will not change for states that are phonon replicas of a single excitonic state. An exciting prospect here is to use the fluence and time dependence of the experimental dephasing rates to estimate the interaction potential and subsequently the scattering length scales, in the same spirit of the phase space filling models proposed by Schmitt-Rink and coworkers~\cite{schmitt1985theory}. This however demands a more formalistic expansion of $V_0$ beyond Born approximation and considering the exact model for the exciton polaron. 

To conclude, reliable estimation of the homogeneous linewidths and their dependencies on various external perturbations is crucial in quantifying exciton-bath interactions. In this perspective article, based on contemporary experimental data and established theoretical models, we argued that coherent nonlinear spectroscopy is the viable methodology to that purpose. Importantly, linear spectroscopies, while being experimentally simple, are \emph{not} capable of providing accurate estimates of the dephasing rates. Thus, photoluminescence linewidths should not be used for the estimation of electron-phonon coupling parameters of metal halide perovskites and their derivatives.

Coherent nonlinear spectroscopies, on the other hand, not only offer the possibility to extract robust homogenous linewidths, but also  offer unprecedented details into the mechanistics of the many-body interactions in materials. Beyond the photon echo implementation discussed in this perpective, other multi-pulse variations including the two-quantum~\cite{Thouin2018stable} and zero-quantum measurements~\cite{liu2019non} can be unique metrologies of excitonic characteristics. Importantly, when combined with simple yet powerful theoretical models, they can be effective tools to assess the spectral density of the bath~\cite{liu2019simultaneous} and providing insights into the material handles to engineer it.

        

\section{Author Information}

\subsection{Notes}

The authors declare no competing interests.

\subsection{Biographies}

Ajay Ram Srimath Kandada completed his Ph.D.\ in Physics at Politecnico di Milano (Italy) in 2013 and was a post-doctoral researcher in the Italian Institute of Technology until 2016. He was a Marie Sklodowska Curie global fellow between 2016 and 2019, during which he performed his post-doctoral research at the Universit\'e de Montr\'eal (Canada) and at Georgia Institute of Technology (USA). In 2020, he joined the Department of Physics at Wake Forest University, USA as an Assistant Professor. His research interests include ultrafast optical spectroscopy of excitonic semiconductors and development of quantum optical probes of materials. 

\noindent Hao Li received his Ph.D. in Physical Chemistry from the Wayne State University in 2011 and was a postdoctoral researcher at the Center of Nonlinear Studies (CNLS) in Los Alamos National Laboratory until 2014. He then joined the University of Houston as a postdoctoral fellow and became a research associate since 2019. His research focuses on the light-matter interactions, excited-state electronic structures, and photoexcitation dynamics in semiconductor materials. 

\noindent Eric Bittner obtained his Ph. D. in theoretical chemistry from the
University of Chicago in 1994 and subsequently was an NSF postdoctoral fellow at the University of Texas at Austin and 
Stanford University. In 1997, he was appointed as an Assistant Professor at the University of Houston where he is currently the
Moores Professor of Chemical Physics.  He is a Guggenheim and a Fulbright Fellow, and holds fellowship in the APS and RSC. 
His research focuses upon light-matter interactions, excited state
dynamics, stochastic processes, and condensed matter physics.

\noindent
Carlos Silva earned a Ph.D.\ in Chemical Physics from the University of Minnesota in 1998 and was then Postdoctoral Associate in the Cavendish Laboratory, University of Cambridge. In 2001 he became EPSRC Advanced Research Fellow in the Cavendish Laboratory, and Research Fellow in Darwin College, Cambridge. In 2005, he joined the Universit\'e de Montr\'eal as Assistant Professor, where he held the Canada Research Chair in Organic Semiconductor Materials from 2005 to 2015 and a Universit\'e de Montr\'eal Research Chair from 2014 to 2017. He joined Georgia Tech in 2017, where he is currently Professor with joint appointment in the School of Chemistry and Biochemistry and the School of Physics, and Professor by Courtesy Appointment in the School of Materials Science and Engineering. He is also Honorary Professor in the Department of Applied Physics of the Centro de Investigaci\'on y de Estudios Avanzados del Instituto Polit\'ecnico Nacional (CINVESTAV Unidad M\'erida). He is Fellow of the American Physical Society and the Royal Society of Chemistry. His group focuses on optical and electronic properties of organic and hybrid semiconductor materials, mainly probed by nonlinear ultrafast spectroscopies.

\begin{acknowledgement}

This work was funded by the National Science Foundation (DMR-1904293). CS acknowledges support from the School of Chemistry and Biochemistry and the College of Science at Georgia Tech. 
The work at the University of Houston was funded in
part by the  National Science Foundation (
CHE-2102506,   
DMR-1903785    
) and the Robert A. Welch Foundation (E-1337). 
\end{acknowledgement}

%
%


\providecommand{\latin}[1]{#1}
\makeatletter
\providecommand{\doi}
  {\begingroup\let\do\@makeother\dospecials
  \catcode`\{=1 \catcode`\}=2 \doi@aux}
\providecommand{\doi@aux}[1]{\endgroup\texttt{#1}}
\makeatother
\providecommand*\mcitethebibliography{\thebibliography}
\csname @ifundefined\endcsname{endmcitethebibliography}
  {\let\endmcitethebibliography\endthebibliography}{}

\end{document}